latex

\documentclass[twocolumn,showpacs,preprintnumbers,amsmath,amssymb]{revtex4}
\usepackage{graphicx}

\begin{document}


\title{Smith-Purcell Radiation from Rough Surfaces}

\author{Zh.S. Gevorkian$^{1,2,\ast}$ }
\affiliation {$^{1}$ Institute of Radiophysics and Electronics,Ashtarak-2,0410,Armenia.\\
$^{2}$Yerevan Physics Institute, Alikhanian Brothers St. 2, Yerevan 0036, Armenia. \\
$^{\ast}$ gevork@yerphi.am}



\date{\today}

\begin{abstract}

{\bf Abstract}
\vskip 0.3 truecm

Radiation of a charged particle moving parallel to a inhomogeneous
surface is considered. Within a single formalism periodic and
random gratings are examined. For the periodically inhomogeneous
surface we derive new expressions for the dispersion relation and
the spectral-angular intensity. In particular, for a given
observation direction two wavelengths are emitted instead of one
wavelength of the standard Smith-Purcell effect. For a rough
surface we show that the main contribution to the radiation
intensity is given by surface polaritons induced on the interface
between two media. These polaritons are multiply scattered on the
roughness of surface and convert into real photons. The
spectral-angular intensity is calculated and its dependence on
different parameters is revealed.
\end{abstract}



\maketitle


{\it Introduction.} Smith-Purcell radiation(SP) \cite{SM53} is
originated when a charged particle travels parallel to a plane
with diffraction grating. Recent renewed interest in this problem
is caused by different applications. Among these applications are
 length determination for short electron bunches \cite{DG02},
creation of monochromatic light source in the far infrared region
\cite{SA89,Sh98,UGK98, KS01,AB04,KK05}, {\it etc.}. Various
theoretical models were proposed for describing the SP;
\cite{To60,BD65,BT73,Ku05,Ks05}, for a brief review of recent
theoretical works  see \cite{ABB05,KaPo06}. Most of these models
deal with the periodical grating in the strong scattering regime
(see below). However in many situations the interface over which
the charge travels is rough . As an example one can mention
chamber walls in storage rings. Even the best treated surfaces
contain roughness. Radiation appearing when a charged particle
moves near a rough surface could be useful for beam diagnostics
\cite{ZHS09}. The influence of the surface roughness on the
transition radiation (originating when particle crosses the
interface between two media) was discussed in
\cite{Bag01,ReRos01}. Roughness-induced radiation for a charged
particle sliding over a surface was experimentally observed in
\cite{Fab73}. In the present paper we study radiation emitted due
to  electromagnetic field scattering on  inhomogeneities of
dielectric constant . We will see below that in the weak
scattering regime it is possible to develop a rigorous theory
describing both periodical and random grating within a single
formalism.

 {\it General Relations} The geometry of the problem is shown in Fig.1.

\begin{figure}
\includegraphics[width=8.4cm]{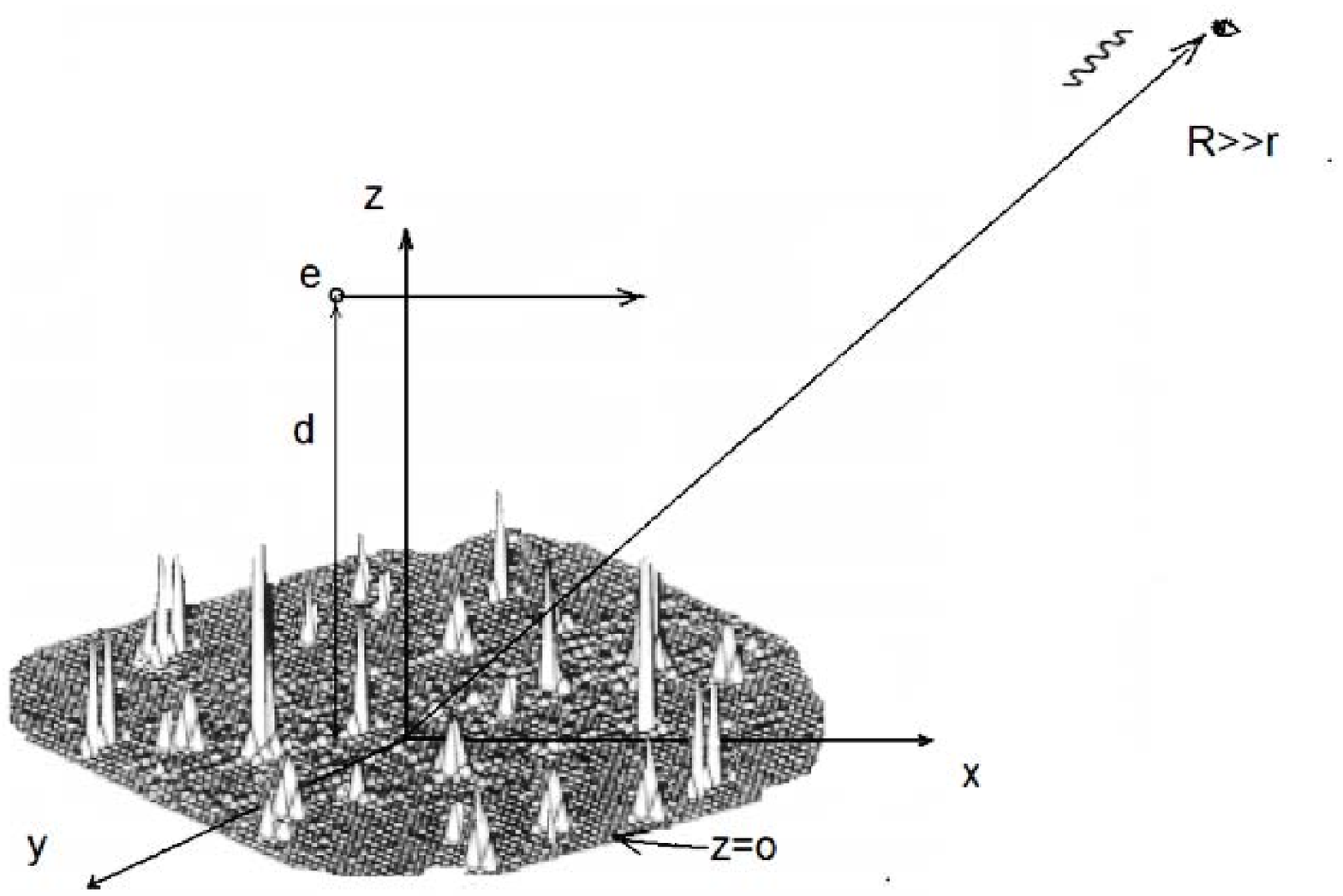}
\caption {Geometry of the problem. A charged particle moves
parallel to the $0x$ axis. Observation point is far away from the
system. }
 \label{fig.1}
\end{figure}

 A charged particle moves uniformly in the vacuum at the
distance $d$ from the plane $z=0$ separating vacuum and isotropic
medium. We are interested in the radiation field far away from the
charge and the interface. The Maxwell equation for the electric
field reads
\begin{equation}
\nabla^2\vec E(\vec r,\omega)-{\rm grad}{\rm div}\vec E(\vec
r,\omega )+\frac{\omega^2}{c^2}\varepsilon(\vec r,\omega)\vec
E(\vec r,\omega)=\vec j(\vec r,\omega)
 \label{Max1}
\end{equation}

where $\vec j$ is the current density related to the charge
\begin{equation}
\vec j(\vec r,\omega)=-\frac{4\pi ie\omega\vec
v}{vc^2}\delta(z-d)\delta(y)e^{i\omega x/v} \label{curr}
\end{equation}

Here $\vec v$ is the velocity of the particle  moving on $0x$
direction and $\varepsilon(\vec r,\omega)$ is the inhomogeneous
dielectric permittivity of the system which for a rough surface
can be chosen in the form
\begin{equation}
\varepsilon(\vec r,\omega)=\Theta[ z-h( x,
y)]+\varepsilon(\omega)\Theta[ h(x,y)-z]. \label{dicons}
\end{equation}

where $\Theta(z)$ is Heaviside$^{\prime}$s unit step function,
$h(x,y)$ is the amplitude of surface roughness. As it follows from
Eq.(\ref{dicons}), the space $z>h( x, y)$ is vacuum while the
space $z<h(x,y)$ is occupied by a medium with isotropic dielectric
constant $\varepsilon(\omega)$. Assuming $h(x,y)$ small and
expanding Eq.(\ref{dicons}), one gets \cite{MM75}
\begin{equation}
\varepsilon(\vec
r,\omega)=\varepsilon_0(z,\omega)+\varepsilon_r(\vec r,\omega)
\label{dicon2}
\end{equation}
where
\begin{eqnarray}
\varepsilon_0(z,\omega)=\left\{ 1,\quad z>0 \atop
\varepsilon(\omega),\quad z<0 \right. \label{dicon3}
\end{eqnarray}
and
\begin{equation}
\varepsilon_r(\vec
r,\omega)=[\varepsilon(\omega)-1]\delta(z)h(x,y). \label{raddic}
\end{equation}
Thus the total $\varepsilon$ is presented as a sum of a regular
part $\varepsilon_0$ and an irregular part $\varepsilon_r$.
 To separate the radiation field we decompose electric field $\vec E=
\vec E_0+\vec E_r$ analogous to Eq.(\ref{dicon2}). Here $\vec E_0$
and $\vec E_r$ are the background and radiation fields,
respectively. They obey the following equations
\begin{eqnarray}
&&\nabla^2\vec E_0(\vec r,\omega)-{\rm grad}{\rm div}\vec E_0(\vec
r,\omega)+
\nonumber \\
&&+\frac{\omega^2}{c^2}\varepsilon_0(z,\omega)\vec E_0(\vec
r,\omega)=\vec j(\vec r,\omega)\label{maxb1}\\
&&\nabla^2\vec E_r(\vec r,\omega)-{\rm grad}{\rm div}\vec E_r(\vec
r,\omega )+\frac{\omega^2}{c^2}\varepsilon_0(z,\omega)\vec
E_r(\vec r,\omega)+\nonumber\\
 &&+\frac{\omega^2}{c^2}\varepsilon_r(\vec r,\omega)\vec E_r(\vec
r,\omega)=-\frac{\omega^2}{c^2}\varepsilon_r(\vec r ,\omega)\vec
E_0(\vec r,\omega)
 \label{maxbacrad}
\end{eqnarray}
Note that although the term $\varepsilon_r E_r$in Eq.(\ref{maxbacrad}) is small one should
 keep it because it causes multiple scattering of electromagnetic field.
 We will see that multiple scattering effects are very important in radiation from rough surface. Multiple scattering effects in SP radiation for a cluster of dielectric particles were discussed in \cite{diabajo}.
At large distances from the system the electromagnetic field can
be treated as a plane wave in which electric and magnetic fields
equal to each other. Therefore the intensity of radiation at the
frequencies $[\omega$,$\omega+d\omega]$ and at solid angles
$[\Omega$,$\Omega+d\Omega]$ can be determined as follows
\begin{equation}
dI(\omega,\vec n)=\frac{c}{2}|\vec E_r(\vec R)|^2R^2d\Omega
d\omega,
 \label{intin}
\end{equation}
where $\vec n$ is unit vector on the direction of observation
point $\vec R$, $\Omega$ is the corresponding solid angle, see
Fig.1 and also \cite{ABB05}. As usual at large distances $|\vec
E_r(\vec R)|^2$ behaves as $1/R^2$ therefore intensity does not
depend on $R$. The expression Eq.(\ref{intin}) should be averaged
over the realizations of random roughness $h(x,y)$. For this
reason it is convenient to introduce the Green's functions of
Eqs.(\ref{maxb1},\ref{maxbacrad})
\begin{eqnarray}
&&\left[\varepsilon_0(z,\omega)\frac{\omega^2}{c^2}\delta_{\lambda\mu}-\frac{\partial^2}{\partial
r_{\lambda}\partial r_{\mu}}+\right.\nonumber\\
&&\left.+\delta_{\lambda\mu}\nabla^2\right]G^0_{\mu\nu}(\vec
r,\vec r^{\prime},\omega)= \delta_{\lambda\nu}\delta(\vec r-\vec
r^{\prime})\label{bare}\\
&&\left[\varepsilon_0(z,\omega)\frac{\omega^2}{c^2}\delta_{\lambda\mu}-\frac{\partial^2}{\partial
r_{\lambda}\partial r_{\mu}}+\delta_{\lambda\mu}\nabla^2+\right.\nonumber\\
&&\left.+\varepsilon_r(\vec r
,\omega)\frac{\omega^2}{c^2}\delta_{\lambda\mu}\right]G{\mu\nu}(\vec
r,\vec r^{\prime},\omega)= \delta_{\lambda\nu}\delta(\vec r-\vec
r^{\prime}). \label{grfun1}
\end{eqnarray}
In Eqs.(\ref{bare},\ref{grfun1}) a summation over the repeated
indexes is supposed. Solutions of inhomogeneous
Eqs.(\ref{maxb1},\ref{maxbacrad}) can be expressed through the
Green's functions Eqs.(\ref{bare},\ref{grfun1}). Using
Eqs.(\ref{maxb1},\ref{maxbacrad}) and (\ref{bare},\ref{grfun1})
one can represent the averaged radiation intensity tensor
$<I_{ij}(\vec R)>=<E_{ri}(\vec
 R)E^*_{rj}(\vec R)>$ in the form
\begin{eqnarray}
<I_{ij}(\vec R)>=\frac{\omega^4}{c^4}\int d\vec r d\vec r^{\prime}
 <G_{i\mu}(\vec R,\vec r)\varepsilon _r(\vec
r)\nonumber\\
G_{\nu j}^{*}(\vec r^{\prime},\vec R)\varepsilon
_r(\vec r^{\prime})> E_{0\mu}(\vec r)E_{0\nu}^{*}(\vec r^{\prime})
\label{radten}
\end{eqnarray}
where the background electric field $E_{0\mu}(\vec r)$ is
expressed through the bare Green's function
\begin{equation}
E_{0\mu}(\vec r)=\int d\vec r_1 G^0_{\mu\lambda}(\vec r,\vec
r_1)j_{\lambda}(\vec r_1) \label{bacfi1}
\end{equation}

Here $<...>$ means  averaging over the surface random profile
$h(x,y)$. Note that in the original Smith-Purcell experiment
\cite{SM53}, as well as in subsequent works on SP a periodical
grating in one direction is used. In this case $h(x,y)\equiv h(x)$
is  some periodical function of one coordinate. In the present
paper within a single approach we consider both periodical and
random gratings. In the random case
 we suppose that $h$ is  a gaussian stochastic process characterized by
two parameters
\begin{eqnarray}
<h(\vec\rho)>=0  \\\nonumber
 <h(\vec\rho_1)h(\vec\rho_2)>=\delta^2W(|\vec \rho_1-\vec\rho_2|)
\label{gauss}
 \end{eqnarray}
where $\vec \rho$ is the two dimensional vector in the $xy$ plane,
$\delta^2=<h^2(\vec\rho)>$ is the average deviation of surface
from the plane $z=0$. Correlation function $W$ is characterized by
a correlation length $\sigma$ at which it is essentially
decreased.

The Maxwell equations for electric fields
Eq.(\ref{maxb1},\ref{maxbacrad}) and Green's functions
Eq.(\ref{bare},\ref{grfun1})should be amended by  boundary
conditions. As usual, it is required that tangential components of
electric field be continuous across the plane $z=0$. The exact
field, of course, will satisfy the boundary conditions across the
surface $z=h(x,y)$ rather than the plane. However this
approximation seems reasonable for small roughness $\lambda\gg
\delta$ and is widely used in the literature. The Green's function
$G_{\mu\nu}(\vec r,\vec r^{\prime},\omega)$, when considered a
function of $z$ for fixed $z^{\prime}$ satisfies the same boundary
condition as the $\mu th$ Cartesian component of electric field.

{\it Green's Functions.} The equation for bare Green's function
Eq.(\ref{bare}) with correct boundary conditions for arbitrary
$\varepsilon(\omega)$ was solved in \cite{MM75}. To obtain
radiation intensity in vacuum we will need Green's functions in
the half space $z>0$. In order to simplify the problem we will
consider the case when isotropic medium is a metal with very large
negative dielectric constant $|\varepsilon(\omega)|\gg 1$. Using
expressions for Green's functions from \cite{MM75} we find the
following basic components
\begin{eqnarray}
G_{zz}^0(\vec p|0,z)=G_{zz}^0(\vec
p|z,0)=\frac{ip^2}{k^2}\frac{\varepsilon(\omega)e^{iqz}}{k_1-\varepsilon(\omega)q}\nonumber \\
 G_{xz}^0(\vec p|z,0)=-G_{zx}^0(\vec
p|0,z)=-\frac{ip_x}{k^2}\frac{\varepsilon(\omega)qe^{iqz}}{k_1-\varepsilon(\omega)q}
\label{grfun2}
\end{eqnarray}
where $G_{ij}^0(\vec p|z,z^{\prime})$ is the two-dimensional
Fourier transform of $G_{ij}^0(\vec r,\vec r^{\prime})$ and $z>0$.
In the coordinate representation
\begin{equation}
G_{ij}^0(\vec r,\vec r^{\prime})=\int\frac {d\vec
p}{(2\pi)^2}e^{i\vec p(\vec \rho-\vec \rho^{\prime})}G_{ij}^0(\vec
p|z,z^{\prime}) \label{fourier}
\end{equation}

Here $\vec p$ and $\vec \rho$ are two-dimensional vectors with
Cartesian components $p_x,p_y,0$ and $x,y,0$. Also $k=\omega/c$,
$k_1$ and $q$ are determined as follows:

\begin{eqnarray}
q=\left\{ \sqrt{k^2-p^2},\quad k^2>p^2 \atop i\sqrt{p^2-k^2},\quad
k^2<p^2 \right. \\
k_1=-(\varepsilon(\omega)k^2-p^2)^{1/2}
 \label{wavenum}
\end{eqnarray}
In Eq.(\ref{wavenum}) a branch cut for the square root along the
negative real axis is assumed \cite{MM75}. Other components of
Green's function are small over the parameter $1/|\varepsilon|$.
 To determine radiation intensity we will need asymptotics of
Green's functions at large distances. Substituting
Eq.(\ref{grfun2}) into Eq.(\ref{fourier}), one finds
\begin{eqnarray}
&&G_{zz}^0(\vec R,\vec
\rho,0)\approx\frac{1}{2\pi\sqrt{2}R}\left[n_z\sqrt{n_{\rho}}\cos\left(
k(R-\vec n_{\rho}\vec \rho)-\frac{\pi}{4}\right) +\right.\nonumber\\
&&\left.+\frac{n_z}{\sqrt{n_{\rho}}}\cos\left(k(R-\vec
n_{\rho}\vec
\rho)+\frac{\pi}{4}\right)\right]+\nonumber\\
&&+\frac{i}{2\pi\sqrt{2}R}\left[\sqrt{n_{\rho}}\cos\left(k(R-\vec
n_{\rho}\vec\rho)+\frac{\pi}{4}\right)-\right.\nonumber\\
&&\left.-\frac{1}{\sqrt{n_{\rho}}}\cos\left(
k(R-\vec n_{\rho}\vec \rho)-\frac{\pi}{4}\right)\right]\nonumber\\
&&G_{xz}^0(\vec R,\vec \rho,0)=-G_{zx}^0(\vec \rho,0,\vec
R)\approx\nonumber\\
&&\frac{1}{2\pi\sqrt{2}R}\left[n_x\sqrt{n_{\rho}}\sin\left(
k(R-\vec n_{\rho}\vec \rho)+\frac{\pi}{4}\right) +\right.\nonumber\\
&&\left.+\frac{n_x}{\sqrt{n_{\rho}}}\sin\left(k(R-\vec
n_{\rho}\vec
\rho)-\frac{\pi}{4}\right)\right]+\nonumber\\
&&+\frac{i}{2\pi\sqrt{2}R}\left[\sqrt{n_{\rho}}n_xn_z\sin\left(k(R-\vec
n_{\rho}\vec\rho)-\frac{\pi}{4}\right)-\right.\nonumber\\
&&\left.-\frac{n_zn_x}{\sqrt{n_{\rho}}}\sin\left( k(R-\vec
n_{\rho}\vec \rho)+\frac{\pi}{4}\right)\right]
 \label{asym}
\end{eqnarray}
 where $\vec n$ is the unit vector on
the direction of the observation point $\vec R=\vec n R$,
$n_{x,z}$ and $n_{\rho}=\sqrt{n_x^2+n_y^2}$ are it's corresponding
components. When obtaining Eq.(\ref{asym}) we use asymptotics of
Bessel functions for large argument \cite{RG65}. Eqs.(\ref{asym})
are correct provided that $kR\gg 1$, $R_{\rho}\gg \rho$ and we use
approximate equation $|\vec R-\vec r|\approx R-\vec n\vec r$.

{\it Radiation Intensity.} Spectral-angular radiation intensity
Eq.(\ref{radten})can be represented as a sum of two contributions,
$I(\vec R,\omega)=I^0(\vec R,\omega)+I^D(\vec R,\omega)$, where
$I_0$ and $I_D$ are single scattering and diffusive contributions,
respectively \cite{Gev98}. First consider the single scattering
contribution to the radiation intensity. Substituting the Green's
functions in  Eq.(\ref{radten}) by the bare ones, we obtain
\begin{eqnarray}
I_{ij}^0(\vec R)=(\varepsilon-1)^2\delta^2k^4\int d\vec \rho d\vec
\rho^{\prime}G^0_{iz}(\vec R,\vec
\rho,0)\nonumber\\
G^{*0}_{zj}(\vec \rho^{\prime},0,\vec R) W(|\vec\rho-\vec
\rho^{\prime}|)E_{0z}(\vec \rho,0)E^*_{0z}(\vec \rho^{\prime},0)
\label{sincon}
\end{eqnarray}
where $(ij)\equiv(xz)$. The background electric field in the limit
$|\varepsilon(\omega)|\gg 1$ can be found from Eqs.(\ref{bacfi1}),
(\ref{curr}) and (\ref{grfun2})
\begin{equation}
E_{0z}(\vec
\rho,0)=-\frac{4ee^{ik_0x}}{v}\frac{dk_0}{\gamma\sqrt{y^2+d^2}}K_1(\frac{k_0\sqrt{y^2+d^2}}{\gamma})
\label{bacfil2}
\end{equation}
where $k_0=\omega/v$, $\gamma=(1-v^2/c^2)^{-1/2}$ is the Lorentz
factor of the particle and $K_1$ is the first order Macdonald
function. As it follows from Eq.(\ref{bacfil2}) the background
electric field and correspondingly radiation intensity is
exponentially small when $\omega d/v\gamma\gg 1$, see also \cite{WW95}.
One can expect essential intensity provided that $\omega d/v\gamma\ll
1$. Far away from the system at the observation point one can use
asymptotic expressions for Green's functions Eq.(\ref{asym}).
Substituting Eqs.(\ref{asym}) and (\ref{bacfil2}) into
Eq.(\ref{sincon}), for the spectral-angular radiation intensity
$I(\omega,\Omega)=cR^2I_{ii}(\vec R)/2$, one obtains
\begin{equation}
I^0(\omega,\Omega)=\frac{e^2}{c\beta^2}\frac{gL_x(1-n_x^2)(1+n_z^2)(1+n_{\rho}^2)}{16\pi
n_{\rho}d} \label{sincon2}
\end{equation}
where $L_x$ is the system size in the $x$ direction,
$g=(\varepsilon-1)^2\delta^2\sigma^2k^4$ and $\beta=v/c$. When
obtaining Eq.(\ref{sincon2}) we neglect strongly oscillating terms
in the limit $kR\gg 1$ and suppose that $W(\vec \rho-\vec
\rho^{\prime})\equiv \sigma^2\delta(\vec \rho-\vec
\rho^{\prime})$. Beside that we  substitute the Macdonald function
by it's asymptotics for small argument assuming that
$k_0d/\gamma\ll 1$. In the opposite limit as was mentioned above
radiation intensity is negligible. The components of unit vector
$\vec n$ are determined through the polar $\theta$ and azimuthal
$\phi$ angles of observation direction: $n_z=\cos\theta,
n_{\rho}=\sin\theta, n_x=\sin\theta \sin\phi$. We consider
radiation into the half-space $z>0$ (vacuum) which means
$\theta<\pi/2$. Note that the coupling constant
$g=k^4(\varepsilon-1)^2\delta^2\sigma^2$ in Eq.(\ref{sincon2})is a
dimensionless parameter. From the condition $R_{\rho}\gg \rho$,
one obtains a restriction on angles $\sin\theta\gg L/R$, where $L$
is a characteristic size of the system. To avoid misunderstanding note that
$1/\beta^2$ dependence of radiation intensity Eq.(\ref{sincon2}) is correct
in an intermediate regime for not very low velocities $\omega d/v\gamma\ll 1$. When $\beta\to 0$,as was mentioned above, radiation disappears.

Note that the background field $\vec E_0$ can originate radiation without any roughness provided that Cherenkov condition $v^2\varepsilon>c^2$ is fulfilled.
Cherenkov radiation is possible for dielectric surfaces with positive large $\varepsilon$. For metallic surfaces in the optical region we are interested in the present paper dielectric constant is negative and Cherenkov radiation is absent.

{\it Periodical case.} Analogously one can consider the case when
surface grating is a periodical function. For simplicity we will
assume that $h(\vec \rho)\equiv \delta \sin2\pi x/b$, where $b$ is
the period of grating. Substituting   $W(|\vec \rho-\vec
\rho^{\prime}|)$ by $\sin2\pi x/b\sin2\pi x^{\prime}/b$ in
Eq.(\ref{sincon}) and using Eq.(\ref{asym}), after integration ,
for spectral-angular radiation intensity, one has
\begin{eqnarray}
I_{SP}(\vec n,
\omega)=\frac{e^2}{c\beta^2}\frac{g_1(1+n_z^2)(1-n_x^2)(1+n_{\rho}^2)L_x}{8\pi
n_{\rho}}\nonumber\\
\left[\delta(kn_x+k_0-\frac{2\pi}{b})+\delta(k_0-kn_x-\frac{2\pi}{b})\right]
F(kn_y) \label{SmPur}
\end{eqnarray}

where $g_1=k^4(\varepsilon-1)^2\delta^2$ and $F$ is determined as
follows
\begin{equation}
F(kn_y)=\left|\frac{dk_0}{\gamma}\int_0^{\infty}dy\frac{K_1(\frac{k_0\sqrt{y^2+d^2}}{\gamma})}
{\sqrt{y^2+d^2}}e^{ikn_y}\right|^2 \label{imp}
\end{equation}
When obtaining Eq.(\ref{SmPur}) we keep only the terms
proportional to $L_x$. In the most interesting case $n_y\sim 0$
and $dk_0/\gamma\ll 1$, substituting $K_1$ by its asymptotic
expression, one finds from Eq.(\ref{imp}),$F(0)\sim \pi^2/4$. As
follows from Eq.(\ref{SmPur}), because of the $\delta$ functions,
for a given observation direction, only two discrete wavelengths
are emitted
\begin{equation}
\lambda_{\pm}=b(\frac{1}{\beta}\pm n_x) \label{disp}
\end{equation}
This is a generalization of well known Smith-Purcell dispersion
relation \cite{SM53} to the weak scattering (see below) case.
 For an arbitrary periodical grating one can expand the surface
profile $h(x)$ into Fourier series and for each term one can
obtain analogous dispersion relation with   $b$  substituted by
$b/m$, where $m$ is the diffraction order. Note that the
dispersion relation Eq.(\ref{disp}) and the spectral-angular
radiation intensity Eq.(\ref{SmPur}) differ from  reported
earlier. The reason of those differences are following. First we
are considering weak scattering regime instead of strong one
considered in above mentioned papers. Our theory is applicable
provided that $(\varepsilon(\omega)-1)^2 \delta^2/\lambda^2 \ll 1$
although $| \varepsilon(\omega)|\gg 1$. Probably this regime was
realized in the experiment on SP radiation in the optical region
for shallow gratings \cite{Ku02}. As mentioned in \cite{Ku02}
traditional formula of SP radiation is failed to explain the
results of experiment in the shallow grating case. Second reason
is the boundary conditions. As follows from
Eqs.(\ref{maxb1},\ref{maxbacrad}) Maxwell equations for $\vec E_0$
and $\vec E_r$ contain the same disruptive function
$\varepsilon_0(z,\omega)$. Therefore both of them should satisfy
the same boundary conditions at $z=0$. In our consideration this
goal is achieved automatically because the Green's functions
Eqs.(\ref{grfun2},\ref{asym}) satisfy the correct boundary
conditions \cite{MM75}. In contrary, in traditional consideration
\cite{BT73}, only the total field $\vec E_0+\vec E_r$ and not they
separately, satisfy the boundary conditions at $z=0$. Probably this
difference leads to different dispersion relation Eq.(\ref{disp}).

 {\it Diffusive Contribution, Surface
Polaritons.} Using Eq.(\ref{radten}) one finds diffusive
contribution to the radiation intensity in the form
\begin{eqnarray}
I^D_{ij}(\vec R)=g\int d\vec \rho_1 d\vec \rho_2 d\vec \rho
G_{im}(\vec R,\vec \rho_1,0)G_{hz}(\vec \rho_2,\vec
\rho)\nonumber\\
P_{mnhs}(\vec \rho_1-\vec \rho_2)
G_{zs}^{*}(\vec \rho,\vec \rho_2)G_{nj}^{*}(0,\vec \rho_1,\vec
R)|E_{0z}(\vec \rho,0)|^2 \label{diffcon}
\end{eqnarray}
where $G_{ij}(\vec \rho_2,\vec \rho_1)\equiv G_{ij}(\vec
\rho_2,0,\vec \rho_1,0)$,and where diffusive propagator $P$ is
determined by the sum of ladder diagrams see Fig.2. and
\cite{AGD63}.

\begin{figure}
\includegraphics[width=8.4cm]{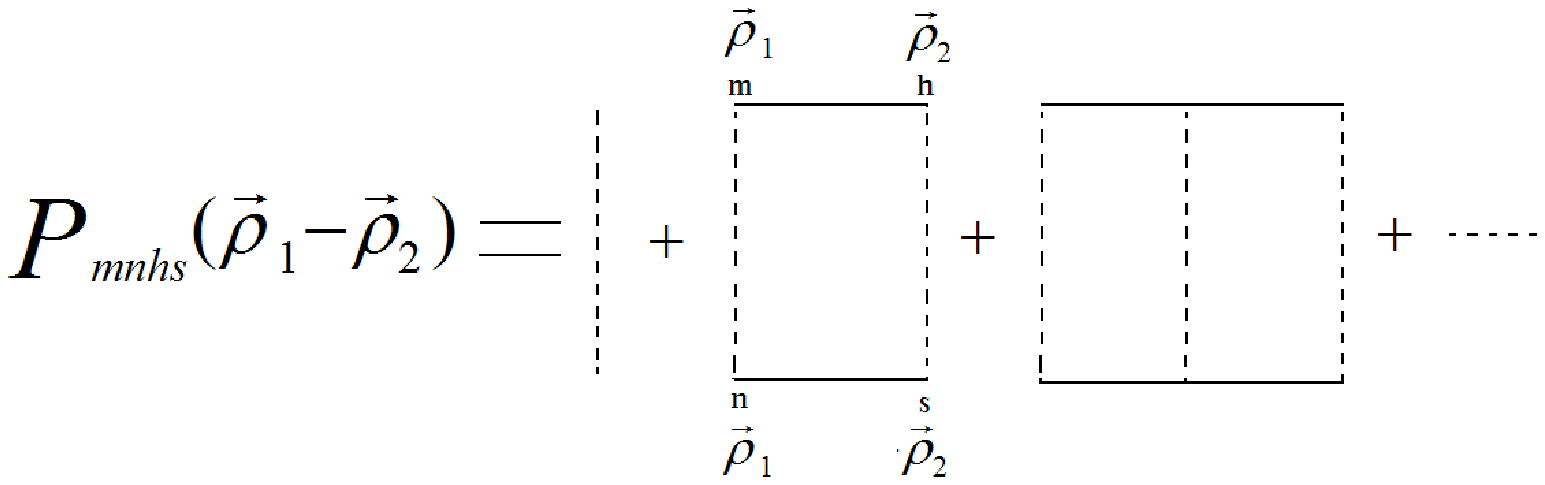}
\caption {Dashed line  is the correlation function of roughness
$g\delta(\vec\rho_1-\vec\rho_2)$ and the solid line is the
averaged over the randomness two-dimensional Green's function of
surface polariton }
 \label{fig.2}
\end{figure}

All integrations over $z$ coordinates make them equal to $0$
because of $\delta(z)$ in fluctuation part of dielectric constant
Eq.(\ref{raddic}).  Averaged two-dimensional surface polariton
Green's function \cite{MM85} satisfies the Dyson equation
\begin{equation}
G_{\mu\nu}(\vec p)=G^0_{\mu\nu}(\vec p)+gG^0_{\mu m}(\vec
p)\int\frac{d\vec p_1}{(2\pi)^2}G^0_{mn}(\vec p_1)G_{n\nu}(\vec p)
\label{Dyson}
\end{equation}

Remind that $G{\mu\nu}(\vec p)\equiv G_{\mu\nu}(\vec p|0,0)$,see
Eq.(\ref{grfun2}). Bare Green's functions are determined by
Eq.(\ref{grfun2}). In further we will interested in the behavior
of Green's function close to the pole. These values give the main 
contribution in the limit  $g\to 0$.
As it follows from Eq.(\ref{grfun2}) two-dimensional Green's
functions of surface polariton has a pole at
$p^2=k^2\varepsilon/(\varepsilon+1)$, see \cite{MM85}. The
corresponding velocity of a surface polariton is equal to
$c\sqrt{(\varepsilon+1)/\varepsilon}<c$. Remind that we consider
the case when $\varepsilon \ll -1$. When electron velocity becomes
equal to this velocity a superradiant emission is possible
provided that the grating is periodical \cite{AB04,UGK98}. Close
to the pole and for large negative $|\varepsilon(\omega)|\gg 1$
the Green's functions of surface polariton can be represented in
the form
\begin{eqnarray}
G_{zz}^0(p)\simeq
\frac{-k}{\sqrt{-\varepsilon_1(\omega)}}\frac{1}{k^2-p^2-i\alpha}\nonumber \\
G_{zx}^0(\vec p)\simeq
\frac{ip_x}{\sqrt{-\varepsilon_1(\omega)}}\frac{1}{k^2-p^2-i\alpha}
\label{pole}
\end{eqnarray}
where $\alpha=k^2\varepsilon_2/\varepsilon_1^2$,
$\varepsilon=\varepsilon_1+i\varepsilon_2$ and
$\varepsilon_2\ll|\varepsilon_1 |$. In Eq.(\ref{pole}) $\alpha$
describes the damping of the surface polariton on the flat surface
due to the inelastic processes in the medium, i.e.
$\varepsilon_2(\omega)$. It follows from Eq.(\ref{pole}) that
$\int d\vec p \vec p  G_{zx}^0(\vec p)\equiv 0$. Therefore only
$G^0_{zz}$ gives contribution to the integral in Eq.(\ref{Dyson}).
Solving Dyson equation Eq.(\ref{Dyson}) one can represent the
averaged Green's functions in the form
\begin{eqnarray}
G_{zz}(p)\simeq
\frac{-k}{\sqrt{-\varepsilon_1(\omega)}}\frac{1}{k^2-p^2-i\Lambda}\nonumber \\
G_{zx}(\vec p)\simeq
\frac{ip_x}{\sqrt{-\varepsilon_1(\omega)}}\frac{1}{k^2-p^2-i\Lambda}
\label{ave}
\end{eqnarray}

where $\Lambda=\int \frac{d\vec p}{(2\pi)^2}Im
G_{zz}^0(p)=gk^2/4\varepsilon_1(\omega)$. Real part of the
integral leads to renormalization of the parameters and does not
play any role. Integral is calculated in the limit
$\varepsilon_2\to 0$. $\Lambda$ describes the damping of the
surface polariton by its roughness-induced conversion into
radiative modes \cite{MM85}. It is convenient also to introduce
the polariton mean free path on the rough surface $l=k/\Lambda$.
Note that the neglected terms in diagram expansion are small on
parameter $\lambda/l\ll 1$ \cite{AGD63}.

 Using Eqs.(\ref{ave}) one sees that the main
contribution to the diffusive radiation intensity
Eq.(\ref{diffcon}) give the term proportional to $P_{zzzz}$ which
contains a diffusive pole at small momentums. Summing the ladder
diagrams in Fig.2 one finds a Bethe-Salpeter equation for
diffusive propagator $P(K)\equiv P_{zzzz}(K)$
\begin{equation}
P(K)=g^2f(K)+gf(K)P(K)
\label{BSal}
\end{equation}
where
\begin{equation}
f(K)=\int\frac{d\vec p}{(2\pi)^2}G(p)G^{*}(|\vec p-\vec K|)
\label{ffunc}
\end{equation}
Here $G(p)\equiv G_{zz}(p)$. Using Eq.(\ref{ave}) and calculating
the integral in Eq.(\ref{ffunc}) in the limit $g\to 0$, one finds
$P(K)$ at small $Kl\ll 1$
\begin{equation}
P(K)=\frac{2g}{K^2l^2}
\label{prop}
\end{equation}
Substituting Eqs.(\ref{asym}) into Eq.(\ref{diffcon}), for the
diffusive contribution to the radiation intensity, one has
\begin{eqnarray}
&&I^D(\omega,\Omega)=\frac{c(n_z^2+1)(1-n_x^2)(1+n_{\rho}^2)}{64\pi^2n_{\rho}}P(K\to
0)\nonumber\\
&&\int d\vec \rho |E_{0z}(\vec \rho,0)|^2 \label{diffcon2}
\end{eqnarray}

As follows from Eqs.(\ref{prop}) and (\ref{diffcon2}) radiation
intensity diverges. This divergence is caused by the infinite size
of the system, see also \cite{Gev98,Gev06}. If one takes into
account the finite size of the system the minimal momentum will be
of order $\sim 1/L$. As was mentioned above the radiation
intensity is exponentially small provided that $d\gg \gamma/k_0$.
In the opposite limit $d\ll \gamma/ k_0$ substituting $K_1$ by it's
asymptotic expression and integrating Eq.(\ref{diffcon2}), we
finally obtain
\begin{equation}
I^D(\omega,\Omega)=\frac{e^2}{c\beta^2}\frac{g(1+n_z^2)(1-n_x^2)(1+n_{\rho}^2)}{8\pi
n_{\rho}}\frac{L_xL^2}{dl^2} \label{diff3}
\end{equation}

In this consideration weak $l\ll l_{in}$,where
$l_{in}=k/\alpha=\varepsilon_1^2/k\varepsilon_2$ is the inelastic
mean free path of surface polariton, absorption can be taken into
account as follows \cite{And85}. When $L>(ll_{in})^{1/2}$,   $L$
in Eq.(\ref{diff3}) should be substituted by $(ll_{in})^{1/2}$.
Comparing single scattering Eq.(\ref{sincon2}) and diffusive
Eq.(\ref{diff3}) contributions, one has $I^D/I^0\sim L^2/l^2\gg
1$. Therefore the diffusion of surface polaritons is the main
mechanism of radiation. Let us make some numerical estimates for
the optical region. For $Ag$ at $\lambda \sim 4500A^o$
$\varepsilon_1\sim -7.5$ and $\varepsilon_2\sim 0.24$. Taking
$\delta\sim 50A^o$ and $\sigma\sim 1000A^o$ \cite{MM85}, one has
$g\sim 0.68$, $l\sim 7.7\lambda$ and $l_{in}\sim 47.94\lambda$.
Thus the conditions for diffusive mechanism $\lambda\ll l\ll
l_{in},L$ are fulfilled. Evidently, depending on grating
parameters $\delta,\sigma$ emission in other wavelength regions is
possible too.

We have considered multiple scattering effects in
radiation for uncorrelated roughness. However they are very 
important for the periodical as well as correlated grating 
cases too. These cases are more complicate and will be discussed 
elsewhere later. Our result Eq.(\ref{SmPur})for SP radiation 
intensity with only single scattering contribution is correct
in the cases when multiple scattering contribution is 
negligible. Such a situation can occur for the metals with
relatively large absorption when the condition of multiple 
scattering of polaritons $l_{in}\gg l$ is not fulfilled.

\acknowledgments

I am grateful to A.Allahverdyan for helpful discussions. This work
is supported by the ISTC grant A-1602.

\end{document}